\documentclass[useAMS]{mn2e}
\usepackage{epsfig}
\title[Collision rate proxies]{On the reliability of proxies for
globular cluster collision rates}

\author[Maccarone \& Peacock]{Thomas J. Maccarone\\ School
of Physics and Astronomy, University of Southampton, Hampshire SO17
1BJ,United Kingdom\\ \newauthor Mark B. Peacock\\ Department of Physics and Astronomy, Michigan State University, East Lansing MI , USA}

\begin{document}
\def\ltsim{\mathrel{\rlap{\lower 3pt\hbox{$\sim$}}
        \raise 2.0pt\hbox{$<$}}}
\def\gtsim{\mathrel{\rlap{\lower 3pt\hbox{$\sim$}}
        \raise 2.0pt\hbox{$>$}}}

\date{}

\pagerange{\pageref{firstpage}--\pageref{lastpage}} \pubyear{}

\maketitle

\label{firstpage}

\begin{abstract}

A variety of different proxies for the stellar collision rates in
globular clusters is used in the literature, depending on the quality
of data available.  We present comparisons between these proxies and
the full integrated collision rate for different King models.  The
most commonly used proxy, $\Gamma$, defined to be $\rho_0^{3/2}
r_c^2$, where $\rho_0$ is the central cluster density, and $r_c$ is
the core radius based on the 1966 King model, is an accurate
representation of the collision rate from the King model to within
about 25\% for all but the least concentrated globular clusters.  By
integrating over King models with a range of parameters, we show that
$\Gamma_h$, defined to be $\rho_h^{3/2} r_h^2$, where $\rho_h$ is the
average density within the half-light radius, and $r_h$ is the
half-light radius, is only marginally better correlated with the full
King model collision rate than is the cluster luminosity.  The two
galaxies where results of King model fitting have been reported in
detail show a dearth of core-collapsed clusters relative to that seen
in the Milky Way, indicating that the core radii of the most
concentrated clusters are probably slightly overestimated, even with
excellent data.  Recent work has suggested that shallower than linear
relations exist between proxies for $\Gamma$ and the probability that
a cluster will contain an X-ray source; we show that there is a
similarly shallower than linear relationship between $\Gamma$ and
$\Gamma_h$ that can explain the relationship where $\Gamma_h$ is used;
we also show that reasonable measurement errors are likely to produce
a shallower than linear relationship even when $\Gamma$ itself is
used.  We thus conclude that the existing evidence is all consistent
with the idea that X-ray binary formation rates are linearly
proportional to cluster collision rates.  We also find, through
comparison with Gunn-Griffin models (sometimes referred to as
multi-mass King models) suggestive evidence that the retention
fractions of neutron stars in globular clusters may be related to the
present day concentration parameters, which would imply that the most
concentrated clusters today were the most concentrated clusters at the
time of their supernovae.

\end{abstract}


\section{Introduction}

The density of stars in most globular clusters is large enough that
significant rates of collisions and other close interactions between
stars can take place.  Globular clusters are known to have about two
orders of magnitude more bright X-ray binaries per unit stellar mass
than field star populations (see e.g. Clark 1975 for the first
evidence of this effect in the Milky Way; Supper et al. 1997; Angelini
et al. 2001; Sarazin, Irwin \& Bregman 2001 for early indications from
individual nearby galaxies; Kundu et al. 2003 for the first
demonstration of a universal trend from a compilation of nearby
galaxies).  The excess of X-ray binaries in globular clusters is
generally explained as a result of X-ray binary formation through
tidal captures (Fabian, Pringle \& Rees 1975), three-body exchange
encounters (Hills 1976) and direct collisions between stars (Verbunt
1987).

Millisecond radio pulsars are also overabundant in globular clusters
by a significant amount (e.g. Camilo \& Rasio 2005), most likely
because they evolve from low mass X-ray binaries (Alpar et al. 1982).
It is generally believed that significant numbers of stars in other
classes, such as cataclysmic variable stars, and blue straggler stars
may be formed through stellar interactions in globular clusters as
well, and considerable debate continues about whether the horizontal
branch morphology of globular clusters is affected by stellar
interactions (e.g. Buonanno et al. 1997; Peacock et al. 2011).

A single tight binary may have a larger binding energy than all the
single stars in a globular cluster put together.  Furthermore, X-ray
sources are the only traces of the dynamical properties of globular
clusters which can currently be well observed at distances of the
Virgo Cluster.  Understanding the formation and evolution of these
binary stars is thus important both in its own right, and for
understanding the evolution of globular clusters in general.

Unfortunately, the Milky Way lacks large enough numbers of bright
(i.e. $L_X>10^{35}$ ergs/sec) X-ray binaries for real statistical
samples to be made.  Only 12 of the Milky Way's globular clusters
contain bright X-ray sources, although two of these clusters now show
clear evidence for containing multiple such sources.  Some intuition
can be developed about which classes of clusters are most likely to
contain X-ray sources, but the sample is not large enough to make
rigorous tests of whether these parameters have a statistically
significant correlation with the probability a cluster will contain an
X-ray binary.  For example, the first four globular cluster X-ray
sources motivated Silk \& Arons (1975) to suggest that metal rich
clusters were more likely than metal poor clusters to contain X-ray
binaries.  Bellazzini et al. (1995) established from the combined
sample of Milky Way and M~31 that metallicity was, in fact, an
important parameter, but the correlations between metallicity and
other parameters, such as galactocentric radius still left some
concerns about which was the causal parameter.
 
The use of more distant extragalactic globular clusters, in the large
samples of clusters which can be seen in nearby elliptical galaxies,
can enhance the sizes of samples of X-ray sources dramatically. Only
with Chandra and HST observations of the nearby galaxy NGC~4472 was it
possible to demonstrate clearly that the probability a cluster will
contain an X-ray source is more strongly correlated with cluster
metallicity than with cluster galactocentric radius (Kundu et
al. 2002).  This finding has since been repeated in many other
galaxies (see e.g. the review article of Maccarone \& Knigge 2007 and
references within).

Another important, but harder-to-address question, is whether the
correlation between the stellar interaction rate in a cluster and the
probability it will contain an X-ray source is linear.  The
measurement of stellar interaction rates in extragalactic globular
clusters is observationally extremely challenging.  Only in M~31 are
typical cluster core radii larger than the angular resolution of the
Hubble Space Telescope, but because M~31 is so nearby, very few of its
clusters are close enough to one another that a single HST field of
view contains multiple clusters.  Nonetheless, one can fit King (1966)
models to data, even if the core radius is formally unresolved, but
only with high signal-to-noise and good oversampling and understanding
of the telescope point spread function (e.g. Carlson \& Holtzman
2001).  Peacock et al. (2009) have presented King model fits for about
half of the globular clusters in M~31 using wide-field ground-based
infared observations, and compared these with an updated catalog of
X-ray sources in Peacock et al. (2010a), while Jord\'an et al. (2004)
have presented some results from King model fits to HST observations
of M~87's clusters, and Jord\'an et al. (2007) have presented some
results from HST observations of NGC~5128.

Some attempts have been made to test whether the probability a
globular cluster will contain an X-ray binary is linearly proportional
to the collision rate, based on both King model fits, and other
proxies for the collision rate.  Jord\'an et al. (2004) showed that,
presuming their King model fits were reliable, the probability a
cluster contains an X-ray binary scales with $\Gamma \rho^{-0.4}$, and
argued that this showed that binaries are being destroyed in dense
clusters.  Smits et al. (2006) showed that the scaling between
collision rate and cluster mass for the Milky Way is such that if one
obtained data with large random errors in the core radii, so that the
collision rate estimates were dominated by the effects of cluster
luminosity, one would obtain results like those found in Jord\'an et
al. (2004).  The results of Jord\'an et al. (2004) were based on data
from the ACS Virgo Cluster survey, which has an integration time of
only 750 seconds in the $g$ band filter. It is thus likely that the
core radii fitted from those data have measurement errors large
compared to the dispersion in the values of the core radius, since
only a few percent of the clusters are bright enough that they should
have the S/N of at least 500 needed to obtain core radii accurate to
within a factor of 2 at the distance of the Virgo Cluster with HST
(see, again, Carlson \& Holtzmann 2001 for discussion to the effect
that this signal to noise is needed).

More recently, Sivakoff et al. (2007) used $\Gamma_h$, defined to be
$\rho_h^{3/2} r_h^2$, where $\rho_h$ is the average density within the
half-light radius, and $r_h$ is the half-light radius to estimate
collision rates.  They found that the probability a cluster would host
an X-ray source was proportional to $\Gamma_h^{0.8}$, and interpreted
this result as additional evidence of binary destruction.  Jord\'an et
al. (2007) present results from fitting King models to data from
NGC~5128, the nearest giant elliptical galaxy.  These data are deeper
than the data obtained in the ACS Virgo Cluster Survey, and the galaxy
observed is a factor of almost 5 closer than M87, lending some
confidence that the core radii fitted will be reliable, at least for
many of the clusters.  They, too find a weaker than linear dependence
of probability that an X-ray binary will exist in a cluster on its
estimated collision rate.

The Galactic globular clusters have also been studied to determine
whether the lower luminosity X-ray sources they contain have numbers
which scale with cluster collision rate.  Pooley et al. (2003) used
all sources brighter than $L_X=4\times10^{30}$ ergs/sec, and found
that the number of sources scales with $\Gamma^{0.74\pm0.36}$,
consistent with a linear correlation at the 1-$\sigma$ level.
Additionally, the lowest collision rate clusters may have their rates
of formation of X-ray sources dominated by primoridal systems, rather
than by dynamical formation (e.g. Pooley \& Hut 2006), so even a
statistically significant finding that the source numbers scaled more
slowly than linearly with $\Gamma$ would have multiple possible
interpretations.

The idea that the dependence of the probability a cluster will host an
X-ray binary is weaker than linear because of binary destruction in
the densest clusters would have major implications on the channels by
which X-ray binaries form.  For the X-ray binaries themselves to be
strongly subject to disruption requires that they be soft -- that is
if their binding energies are smaller than the mean kinetic energies
of the cluster's single stars (Heggie 1975).  Alternatively, the LMXBs
could, potentially, be formed from exchange encounters where the
targets are soft binaries, in which case it might be possible for the
LMXBs themselves to be hard binaries, while their formation rate
depends on the number of soft binaries in the cluster.  The hard-soft
boundary in most globular clusters occurs for orbital separations of
about 1 AU -- a size scale on which only highly evolved stars can
overflow their Roche lobes.  Of the Milky Way's globular cluster X-ray
binaries with known orbital periods, the longest period system is
AC~211 in M~15, whose orbital period is about 19 hours.  It thus seems
unlikely that binary destruction plays a major role in regulating the
number of X-ray binaries, but given the small number statistics of
globular cluster X-ray binaries with known orbital periods, and the
importance of the implications for understanding how globular cluster
X-ray binaries form, the suggestion merits further investigation.
What is missing from the present analysis in an understanding of how
the measurement errors in the samples of cluster parameters made to
date, and the systematic offsets between real collision rates and
proxy collision rates, affect the conclusions which can be drawn from
the data.  In this paper, we look into both issues.

\section{King models and collision rates}

Given a King model fit, it is customary to estimate the cluster's
stellar collision rate as $\Gamma \propto \rho_0^{3/2} r_c^2$, as first
suggested by Verbunt \& Hut (1987).  The underlying assumptions of
this calculations are that the cluster is well described by a
single-mass King model (and hence is in virial equilibrium); that the
cluster's core density and core velocity dispersion are constant
within the core; and that collisions take place only in the
core. Further complications may arise if one wishes to consider the
collision rates of neutron stars, as is most appropriate for the bulk
of globular cluster X-ray binaries.  One must then, additionally,
consider the retention rate of neutron stars in the globular cluster.
One should also consider the effects of varying characteristic
velocities of interaction in different clusters.  The importance of
this effect has been considered in a study of the dynamical formation
of X-ray binaries in galactic nuclei (Voss \& Gilfanov 2007), and is
implicitly taken into account in dynamical simulations, but has not
generally been considered in more empirically motivated studies of
globular clusters.  Additionally, the lifetimes of many classes of
accreting binaries can be long compared to the relaxation timescale of
a globular cluster, and it would be more appropriate to use a weighted
integral of the collision rate over time (if such a quantity were
measurable), rather than the actual present collision rate (Fregeau
2008).  Therefore, even if it can be confirmed that the probability a
cluster will host an X-ray binary is not linearly proportional to the
collision rate, there are a variety of mechanisms apart from binary
destruction that may be responsible for such deviations.
Understanding the quantitative details of such deviations may provide
a key target for numerical simulations, so it is important to
understand the systematic observational biases that may come into
attempts to measure the X-ray binary production probability versus
collision rate.

It is straightforward to investigate the effects of the assumptions
that the cluster's core density and core velocity dispersion are
constant, and that collisions take place only in the core.  To do
this, we have computed King models for a range of values of cluster
concentration, and have computed the collision rates in each of these
model clusters.  We integrate the collision rate per unit volume
$\rho^2/\sigma$, where $\rho$ is the cluster density, and $\sigma$ is
the cluster's stellar velocity dispersion.  Pooley et al.  (2003)
performed a similar procedure to estimate the collision rates of Milky
Way clusters, but integrated out only to the half light radius.

\begin{figure}
\epsfig{file=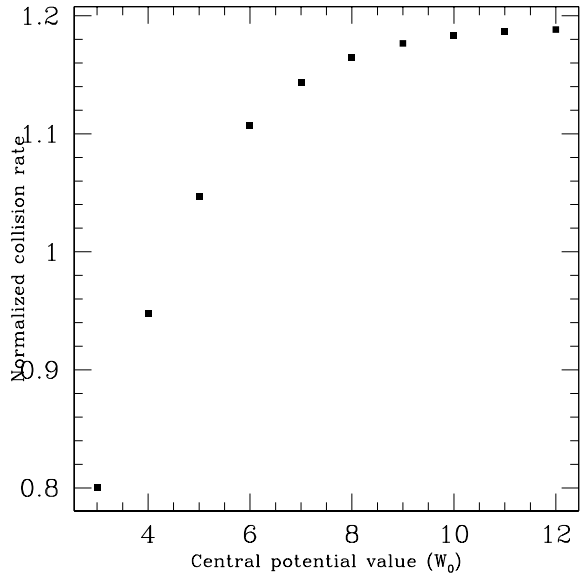,width=8 cm}
\caption{The integrated collision rates from a single-mass King model
plotted versus the value of the central potential $W_0$.  It is clear
that the full integrations are reasonably well-matched by the Verbunt
\& Hut approximation over the full range of values of concentration
see in the Galaxy and M~31, and that the approximation gets better as
one moves to larger values of $W_0$.  The normalization for the
collision rate is by the Verbunt \& Hut formula's value, multiplied in
units that place unity in the middle of the scale.}
\label{integration}
\end{figure}

Following King (1966), we define the cluster concentration as the
ratio of the cluster tidal radius to its core radius. We normalise the
density in units of the central density and the radius in units of the
core radius.  The velocity dispersion is normalized according to
equation 31 of King (1966) -- in units of the velocity dispersion of a
King model with $W=\infty$, $(8\pi G r_c^2 \rho_0/9)^\frac{1}{2}$.
Therefore, a simple integration of the King model's collision rate in
these units yields a collision rate in units of the rate derived by
Verbunt \& Hut (1987) times a constant.  In figure \ref{integration}
we plot the collision rate from the King model versus the cluster
concentration.  As is clear from the figure, for all realistic values
of the concentration parameter, the ratio between the collision rate
integrated from the King model and that from the Verbunt \& Hut
approximation varies by only 50\% over the range from $W_0=3$
(corresponding to the least concentrated clusters in the Galaxy) to
$W_0=12$ (correspondingly to the most concentrated clusters in the
Galaxy) -- the approximation is thus accurate to 25\% (modulo effects
of the deviations between single mass King models and real clusters).
While, for purposes of clarity, we did not plot all the models
computed, most of this variation occurs between $W=3$ and $W=5$, with
the approximation of Verbunt \& Hut clearly getting much worse for the
smallest values of $W_0$, as can be seen in figure \ref{integration}.

An additional moderately serious issue remains, in that using
$\Gamma=\rho_0^{3/2}r_c^2$ implicitly assumes that the stellar
velocity dispersion in the cluster core can be estimated correctly
from the virial theorem, assuming a fixed mass-to-light ratio for all
clusters, which is an approximation in good, but not perfect agreement
with the data.  For this reason, Pooley et al. (2003) used observed
velocity dispersions taken from the compilation of Pryor \& Meylan
(1993) to normalize their collision rates, rather than inferred ones
from the virial theorem.  They find deviations of up to a factor of 10
between $\Gamma$ as defined by Verbunt \& Hut (1987) and the collision
rates they compute.  While it is not stated explicitly by Pooley et
al. (2003), the discrepancies are dominated by the differences between
observed and inferred velocity dispersions -- {\it not} the deviations
from the $\Gamma$ formulation and the more detailed treatment of the
interaction rate used by Pooley et al. (2003).  While in principle,
observed velocity dispersions should be more reliable than model
velocity dispersions, the sample of Pryor \& Meylan (1993) is derived
in an inhomogeneous manner, with some clusters measured using
individual stars, and others using integrated light measurements.  The
sample is additionally not focussed on the cluster cores.  For a
couple of the more concentrated clusters, M~15 and M~70, there are
several $\sigma$ discrepancies between the two methods of measuring
the velocity dispersion observationally.  Therefore, it may be that
the velocity dispersions estimated from a King model and a standard
mass-to-light ratio are actually more reliable than the tabulated
observational data.  This issue certainly bears additional attention
in the future.

\subsection{Effects of using a multi-mass model}
It also stands that a single mass King model cannot describe globular
clusters properly.  For studies of X-ray binary production, it is
clear that there will be mass segregation in clusters, and that the
amount of mass segregation may vary from cluster to cluster, since the
relaxation times of the clusters vary substantially.  For many
Galactic globular clusters, the surface brightness profile cannot be
fit properly with a King model.  Several are elliptical (Harris 1996),
and others show evidence for density cusps in their centers
(e.g. Noyola \& Gebhardt 2006).  The deviations between the real
collision rates and the collision rates estimated assuming a single
mass King model may be substantial, but cannot be estimated through
simple analytic formulae.  Additionally, core collapsed clusters are
not well-described by King models.

To address the issue of mass segregation, we have also tried using
multi-mass models similar to the King model.  We have followed the
prescription of Gunn \& Griffin (1979) for producing these models.  We
have chosen to ignore the effects of anisotropy that are possible
within the Gunn \& Griffin (1979) framework (following the work of
Michie 1963) by setting the anisotropy radius to $\infty$.  In the
computational approach suggested by Gunn \& Griffin (1979), the key
parameters are the central potential of the cluster, which is the same
as that used in the King model, the masses of the different species
(i.e. mass classes), and the central densities in the different
species, denoted as $\alpha_j$ where $j$ is the number of the mass
class.

We start first with a simple case of a two component model.  The
heavier component is taken to have a mass of $1.4 M_\odot$
(i.e. roughly than of a neutron star) while the lighter component is
taken to have a mass of $0.7 M_\odot$ (i.e. roughly that of a turnoff
star).  We then compute the collision rates only between the heavy and
the light component, rather than internally to either component.  We
search over a range in $W_0$ from 3 to 9, with larger values avoided
for computational reasons.  We start off with 3\% of the mass in the
heavy component.  We find that for the whole range of central
potentials, very close to half the heavy stars end up in the core
(using the core radius for the mean stellar mass to define the core).
Adding in an additional component of 0.5 $M_\odot$ stars increases the
fraction of the 1.4 $M_\odot$ objects in the core, but the fraction of
the heavy objects in the core still remains roughly constant over a
fairly wide range in central potential.

In this case, where the ratio of central density in the heavy
component to the total central density is constant, the X-ray binary
formation rate will be well-modeled by the Verbunt \& Hut formula --
the range of rates of integrated collision rates to Verbunt \& Hut
formula estimates spanned in the three mass species case is only about
10\%.  If, on the other hand, the neutron stars represent a constant
fraction of the total cluster mass, rather than of the total core
mass, one would expect an increase in the X-ray binary fraction in the
most concentrated clusters.  This effect can be substantial,
approaching a factor of $\sim10$ between the most and least
concentrated clusters -- if one assume that the same fraction of the
mass in any cluster will be in neutron stars, then the most
concentrated clusters should have $\sim10$ times as many X-ray
binaries per unit collision rate as the least concentrated clusters.

\begin{figure}
\epsfig{file=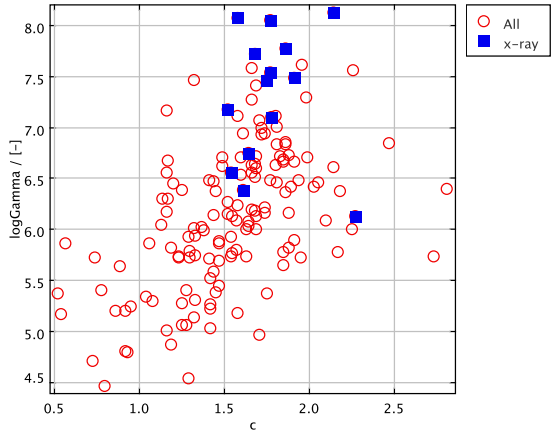,width=8 cm}
\caption{The collision rate $\Gamma$ versus the concentration
parameter [i.e. log$_{10}(r_t/r_c)$] for the M~31 globular clusters.
The filled blue squares are clusters with X-ray sources, while the open red
circles are clusters without X-ray sources.  There is clearly no
preference for clusters with a high concentration parameter at a given
collision rate.}
\label{conc_gamma}
\end{figure}

There is no evidence for such an effect in M~31 (see figure
\ref{conc_gamma}), nor is there any evidence for such an effect in the
Milky Way (for which the analagous plot is similarly devoid of any
conclusive evidence).  If we use the idea that the interaction rate of
neutron stars determines the number of X-ray binaries as a default
assumption, the lack of any obvious effects of $c$ can be taken to
indicate that the neutron star retention fraction is higher in low
concentration clusters than in high concentration clusters.  This
could be the case if, for example, the clusters' structural parameters
have not evolved strongly since the time of the supernovae in which
the neutron stars formed (see for example Figure 15 of Giersz \&
Heggie 2011, which shows that the ratio of core radius to radius
containing 90\% of the stars of a simulated cluster attempting to
match the present day conditions in 47~Tuc changes very slowly over
its lifetime) and the only neutron stars retained were those formed in
the cluster cores, so that the clusters with larger ratios of core
mass to cluster mass have higher retention fractions.  Obviously,
given that the effect has been seen in a sample with fewer than 40
X-ray bright clusters observationally, and in a single set of
dynamical simulations, it would be premature to draw strong
conclusions, but this issue bears further attention, and it certainly
seems reasonable that the neutron star retention fraction may be a
strong function of cluster concentration.

\subsection{Comparison of $\Gamma$ with other tracers of the interaction rate}
\begin{figure}
\epsfig{file=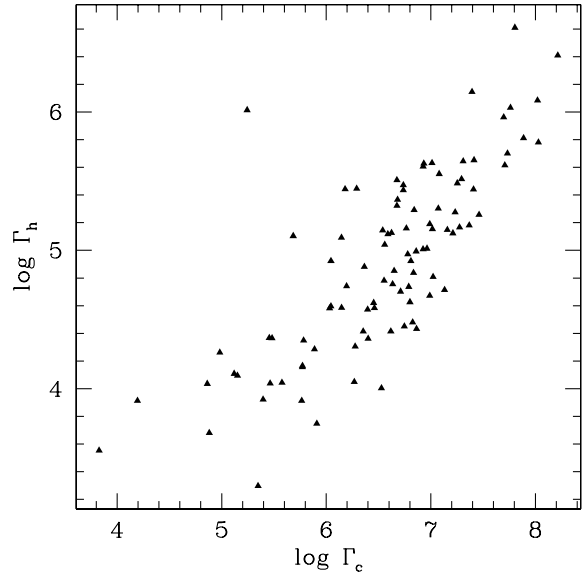,width=8 cm}
\caption{The plot of $\Gamma_h$ versus $\Gamma$ for the 96 Milky Way
clusters where the distance, surface brightness, and luminosity are
given in the Harris catalog. The plot shows the large scatter between
the two different estimators of the stellar collision rate.}
\label{gammah_gammac}
\end{figure}

\begin{figure}
\epsfig{file=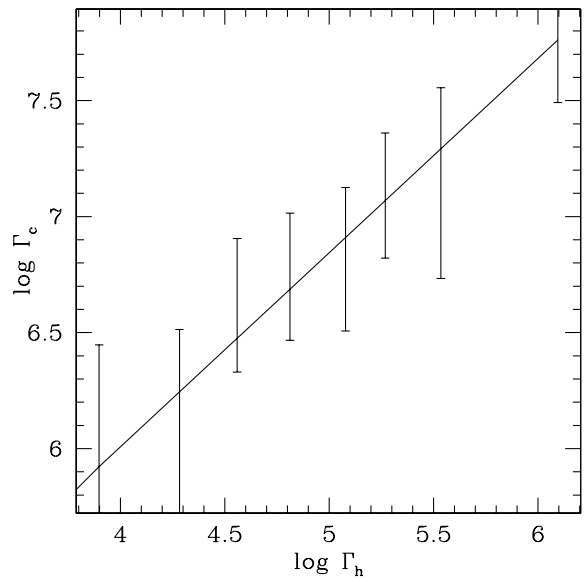,width=8 cm}
\caption{The plot of $\Gamma$ versus $\Gamma_h$ for the 8 bins of 12
Milky Way clusters each where the distance, surface brightness, and
luminosity are given in the Harris catalog. The fitted trend line,
$\Gamma \propto \Gamma_h^{0.83}$ is remarkably similar to the X-ray probability versus $\Gamma_h$ trend found by Sivakoff et al. (2007).}
\label{binned_gammas}
\end{figure}

We can investigate the degree of accuracy of collision rates estimated
through other techniques relative to the collision rate estimated from
the King model rate.  Of these suggested tracers of collision rate,
the easiest to measure is $\Gamma_h$ (Sivakoff et al. 2007).  In order
to assess how useful $\Gamma_h$ is, we can compare $\Gamma_h$ with
$\Gamma$ for the Milky Way's globular clusters from the Harris
catalog.  In figure \ref{gammah_gammac} we plot these two quantities
against one another. Following the methodology in Smits et al. (2006),
which was used to test the correlations expected between collision
rate and mass, we compute $\Gamma_h$ for the 96 Milky Way globular
clusters for which all the relevant quantities are tabulated in the
Harris catalog.  We then sort the clusters into 8 bins of 12 clusters
each, sorted by $\Gamma_h$.  We compute the mean $\Gamma$ in each bin,
as well as the dispersion of values, and fit a power law to the binned
values.  The best fitting power law index is 0.83, very similar to the
exponent of $0.82\pm0.05$ found by Sivakoff et al. (2007) for the
relation between X-ray binary probability and $\Gamma_h$.  A plot of
these binned data is shown in figure \ref{binned_gammas}.  The
$1$-$\sigma$ uncertainty on this exponent is about 0.2 dex, which is
unsurprising given the large scatter within each bin -- it is thus
true that $\Gamma_h$ is formally consistent with being linearly
related to $\Gamma_c$ for the Milky Way sample.  A much larger sample
of clusters would be needed to beat down the uncertainties in the
correlation index that arise due to the variance in the concentration
parameter at any given value of $\Gamma_h$.  Nonetheless, the close
similarity of the two indices, is intriguing, and for non-linear
correlations between $\Gamma_h$ and the sizes of collisionally
produced populations to be taken as indicative of real physics, the
converse should be established -- that the correlation between
$\Gamma_h$ and other measures of the collision rate is linear, or at
least is well-enough established that one can correct for its
deviations from linearity.  

\begin{figure}
\epsfig{file=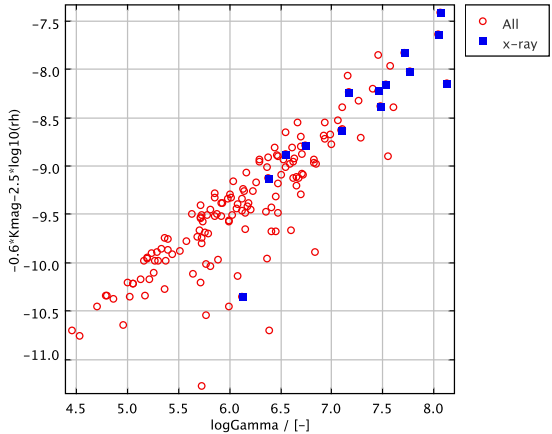,width=8cm}
\caption{The plot of $\Gamma_h$ versus $\Gamma$ for the confirmed M31
clusters from Peacock et al. (2010b).  The filled blue squares
represent the clusters with X-ray sources, while the open red circles
represent the clusters without X-ray sources.  It is clear that the
X-ray clusters skew to the right -- to higher $\Gamma$ for a given
$\Gamma_h$ -- the median X-ray cluster in each block of 0.5 dex in
$Gamma_h$ lies to the right of the median cluster.}
\label{M31_ggh}
\end{figure}

We can also look at the predictive power of the two measures of
collision rate for whether a cluster will have an X-ray source.  This
is done most easily in M~31, since the key parameters for the Galactic
clusters discovered since the New General Catalog was produced are
poorly constrained, and these clusters make up a significant fraction
of the Galactic globular clusters which host X-ray sources.  We
present a plot illustrating that $\Gamma$ is a much better predictor
of whether a cluster will host an X-ray source than $\Gamma_h$ in
figure \ref{M31_ggh}.

\subsubsection{When is $\Gamma_h$ useful and why?}

It is also instructive to compare the relative ability of $\Gamma_h$
and cluster integrated luminosity to predict $\Gamma$ reliably.  Even
if $\Gamma_h$ is not perfectly correlated with $\Gamma$, it may be
adding useful information to a crude understanding of the collision
rates of clusters in galaxies where it is not feasible to obtain
reliable core radii.  There certainly exists a range of distances and
signal-to-noise ratios where it is possible to measure $\Gamma_h$ but
not possible to measure $\Gamma$, so it is interesting to try to
understand in what cases a large sample of measurements of $\Gamma_h$
might be more useful than measurements of the cluster luminosity
alone.

To do this, we take each quantity and make an unweighted fit of a
power law with index 1.0 (i.e. a straight line forced to go through
the origin).  We fit $\Gamma$ as a function of either $\Gamma_h$ or
$L_V$.  We then can compute the variance of the residuals from this
model estimate, in $\Gamma$, so that the variance is independent of
normalization of $\Gamma_h$ or $L_V$.  This indicates that the typical
error in predicting $\Gamma$ from $\Gamma_h$ is 0.58 dex (i.e. a
factor of about 3.8).  The typical error in predicting $\Gamma$ from
cluster luminosity is 0.76 dex (about a factor of 5.8).  It can thus
be seen that the inclusion of half-light radius information, if it is
reliable, adds some, but very little, predictive power, relative to
knowing the cluster luminosity alone.  This is not surprising, given
that half light radius is positively, albeit very weakly correlated
with cluster core radius.

Only for the very most spatially extended clusters in the Milky Way is
there a strong correlation between core radius and half light radius
(see e.g. Djorgovski \& Meylan 1994); see also Figure 3, which shows
plots of half-light radius versus core radius for M~31 [with the data
from the Peacock 2010b catalog], which looks essentially the same in
the plot as the Milky Way in Djorgovski \& Meylan (1994).  It is also
interesting to compare this figure with figure 9 of Sivakoff et
al. (2007), where it is shown that the probability a globular cluster
will contain an X-ray source has a very weak dependence on $r_h$,
except at half light radii larger than about 3 pc -- the same radius
at which the core and half light radii of Milky Way and M~31 clusters
become very strongly correlated and approximately equal to one
another.  Unfortunately, neither plots of $r_c$ versus $r_h$ nor
tables of fitted parameters are presently available for Cen~A and
M~87, so the comparison cannot be extended to those galaxies.
Nonetheless, the data for the Milky Way and M~31 show clearly why
$\Gamma_h$ provides some information about the dynamical state of the
cluster, but not enough that deviations from linearity between
$\Gamma_h$ and probability a cluster containts an X-ray binary can be
taken as strong evidence about what is happening inside a cluster
(e.g. binary destruction).

\begin{figure}
\epsfig{file=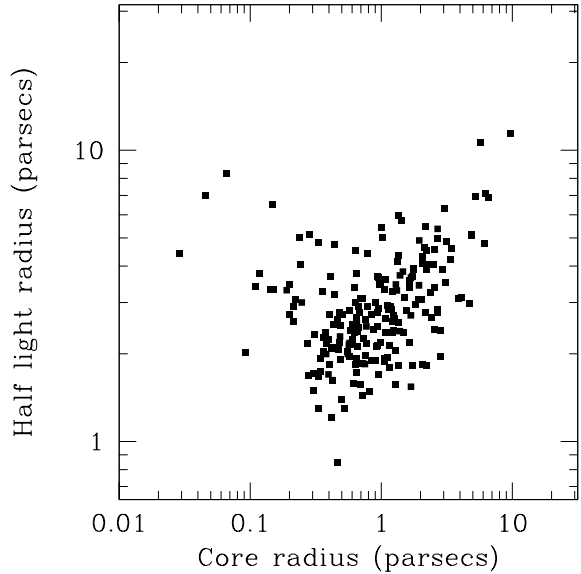,width=8cm}
\caption{The plot of half light radius versus core radius for M31,
using data from Peacock et al. (2010b), restricted to only those
clusters classified as confirmed old clusters in Peacock et
al. (2010b).  The plot is strikingly similar to a similar plot made for
the Milky Way's clusters which appears in Djorgovksi \& Meylan (1994)
as the upper right panel of their Figure 5.}
\end{figure}

\subsection{Reliability of actual estimates of $\Gamma$}
Next, we consider the cases where high quality estimates of the core
radii have been obtained.  At the present time, such measurements have
been done for only two galaxies -- NGC~5128 (Jord\'an et al. 2007) and
M~31 (Peacock et al. 2009).  In both cases, the fraction of clusters
with very high concentration parameters is significantly smaller than
it is in the Milky Way.  In the case of Jord\'an et al. (2007), this
problem may be, in part, due to their decision to excise from their
analyses any clusters with fitted concentration parameters $c$ greater
than 2.5.  Peacock et al. (2009) found several clusters in M~31 which
showed best-fitting concentration parameter values larger than 2.5
among their confirmed clusters.

In the Milky Way, there are 15 known bright X-ray binaries (including
transients), in 12 clusters (see Verbunt \& Lewin 2006 and references
within for the first 13; Altamirano et al. 2009 for the 14th [in
NGC~6440) which was already known to host at least one bright X-ray
binary; and Bordas et al. 2010, Heinke et al. 2010 for the 15th].
Only 31 of the Milky Way's 150 clusters are core collapsed, but the
core collapsed clusters contain 7 of the 14 bright X-ray binaries.
There is only a 1\% chance of this high a fraction of the X-ray
binaries being found in core collapsed clusters at random.  It is thus
important both to count properly the core collapsed clusters, and to
estimate their collision rates correctly to make reliable statements
about specific relations between collision rate and X-ray binary
probability.

We may investigate now what size of errors are needed in order to
produce a deviation from a linear correlation between X-ray binary
hosting probability and collision rate estimator of order the size of
the deviation seen by Jord\'an et al. (2007).  We adopt an approach
intermediate between making a full simulation of all possible
measurement errors, and merely assuming that all values are measured
accurately to arbitrary precision.  We consider both the cases of
additive and multiplicative errors.

As a first step for estimating the effects of measurements errors, we
take the data values from the Harris catalogue, and convert the core
radii into units of parsecs, as is appropriate for comparison with
extragalactic clusters.  First we consider the case of additive
errors.  We add a random number taken from a Gaussian distribution
with varying values of $\sigma$.  If this change gives a cluster a
negative core radius, we set its core radius to be the absolute value
of that number.  We assume that the core luminosity will be fairly
reliably measured, so, as a simplifying assumption, we re-set the core
density such that core luminosity is conserved.  This assumption is
the most favourable one to the idea that the $\Gamma$ measurements are
precise enough to look for deviations from a linear relation between
collision rate and probability of the cluster containing an X-ray
binary.  The key issue we wish to test is that of the effects of
asymmetric error contours that can result even from symmetric errors
in the parameter values for the parameters fitted explicitly.
Assuming a correctly measured core luminosity will produce smaller
devations from linearity and from the expectation value of the
measurement than allowing for radius measurement errors while assuming
a constant core density.  It is thus overly conservative -- serious
problems for interpretations of data may result from errors a bit
smaller than the ones we quote.

We then compute $\Gamma'$ using the standard formula for core
collision rate, but with the values with the simulated measurement
errors rather than the actual values.  We compare the collision rates
for the clusters with the artificial measurement errors added to those
without the artificial measurement errors added, using the same
approach as was used for testing the validity of $\Gamma_h$ -- binning
the data and fitting a power law to the binned data.  We try several
random number seeds, in addition to several values of $\sigma$.  We
find that typically for $\sigma$ of about 0.08 pc, which corresponds
to about 0.08 HST pixels at the distance of Cen A, the additive errors
are large enough to give $\Gamma \propto \Gamma'^{0.85}$.  This value
is about 10\% of the typical core radius value for the clusters in the
Milky Way.  We note that the effects are strongest in the highest
collision rate clusters, which tend to have the smallest core radii.

\begin{figure}
\epsfig{file=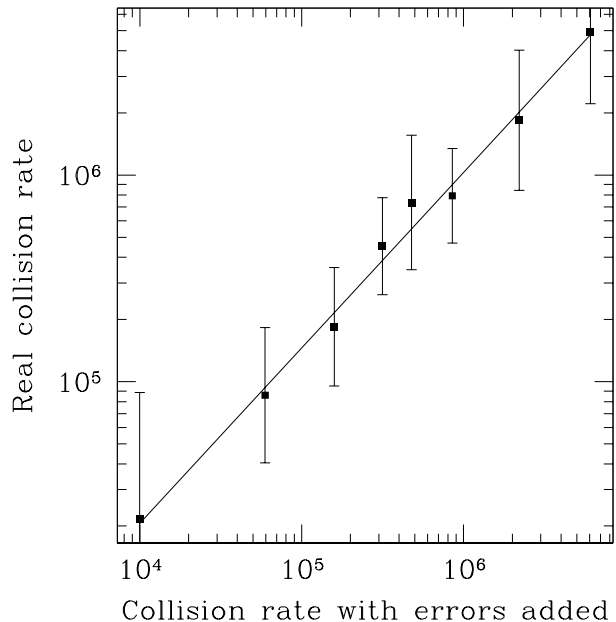,width=8 cm}
\caption{The real collision rates plotted versus the collision rates
with 12\% multiplicative errors added to the core radii while the core
luminosities are held fixed.  The error bars are the scatter in the
real collision rates within each bin.  The line through the data is
one with a slope of 0.85.}
\label{plot_with_errors}
\end{figure}

We can also consider the effects of a multiplicative error on the
cluster core radii.  We perform the same procedure as above, except
that we multiply by a number following a log normal distribution,
rather than adding a number drawn from a normal distribution.  We find
in this case that a multiplicative error with $\sigma$ of 0.12 dex
(i.e. about 30\%) variation from the actual values typically gives a
$\Gamma \propto \Gamma'^{0.85}$ relation (with a 1$\sigma$ uncertainty
of 0.17 dex).  A plot for this case is given in figure
\ref{plot_with_errors}.

In reality, the errors will be correlated both with the
signal-to-noise of the clusters, and with the core radii of the
clusters, so the above results are merely illustrative of the
characteristic magnitudes of errors needed to produce spurious
deviations from linearity.  Jord\'an et al. (2007) cite a detection
limit of $m_{F606W} \sim 22$ for their clusters.  We presume this to
be the $5\sigma$ detection threshold, so that a magnitude of 19.5
corresponds to a 50$\sigma$ detection, but it is not explicitly stated
in Jord\'an et al. (2007) what significance level is implied by the
``detection limit.''  If this is the case, then less than 1/3 of the
clusters are detected at above the $50\sigma$ level, while about half
are detected below the 30$\sigma$ level.  Carlson \& Holtzman (2001)
presented no results for simulation for clusters with signal-to-noise
less than 55, but also presented no results for clusters with core
radii smaller than 0.28 pixels.  They find that the cluster
concentrations can be measured accurately (to within 0.3 dex, or a
factor of 2) 50\% of the time where the core radius is 0.1 pixels and
the signal-to-noise is at least 900, and 50\% of the time for core
radii of 0.28 pixels at signal to noise of at least 100.  It thus
seems likely that the measurement errors could cause the non-linear
relation seen between estimated core collision rate and X-ray binary
hosting probability in NGC~5128, even if the underlying physical
relation is linear.  The situation for the M~87 $\Gamma$ estimates in
Jord\'an et al. (2004) is obviously more severely affects by
measurement errors.

One can also obtain an estimate of typical errors in fit quality by
looking at the variations between the measurements of individual
clusters in Peacock et al. (2009), as that study included a large
number of clusters measured multiple times.  The best fitting core
radii were found to be within 30\% of one another only for clusters
with S/N greater than about 500.  The spatial resolution of the data
of Peacock et al. (2009), using ground-based infrared observations to
estimate core radii of M~31 clusters are only slightly worse than
those the Hubble Space Telescope data taken for Centaurus A, making it
unlikely that the core radii for NGC~5128 clusters are measured to
30\% accuracy with signal-to-noise of 30 or below, as is the case for
the majority of the Cen A clusters.  Additionally, there may be
systematic offsets in the core radius values for NGC~5128, since the
King model fits in that work make use of an analytic approximation to
the point spread function of HST; the point spread function for the
M~31 work was derived directly from measurements of foreground stars,
something possible given the large fields of view used and the
moderate Galactic latitude of M31.

\section{Conclusions}
We have critically evaluated the reliability of various methods for
determining the stellar encounter rates in extragalactic globular
clusters, and their implications for determining whether the
probability a globular cluster will contain a bright X-ray binary
scales linearly with the cluster collision rate.  We show that the
core collision rate of Verbunt \& Hut (1987), $\Gamma = \rho^{3/2}
r_c^2$ gives accurate measurements of the collision rate over a wide
range of cluster parameters, provided that the cluster really is
well-described by a single-mass virialized King model.  We show that
using estimates of the cluster's density within its half-light radius,
plus the half-light radius itself to estimate the cluster's collision
rate give estimates of the collision rates only marginally better than
those which can be obtained by assuming that the collision rate scales
with the cluster mass.  Furthermore, the relationship between
$\Gamma_h$ and $\Gamma$ deviates from a linear relationship in the
same way as the relationship between the measured values of $\Gamma_h$
and the measured values of the probability a cluster will contain an
X-ray binary for Virgo cluster galaxies.  We thus conclude that it is
dangerous to try to infer physical information about how X-ray
binaries are formed by using $\Gamma_h$ -- the deviations from
linearity can easily be explained by the non-linear, and highly
scattered, relation between $\Gamma$ and $\Gamma_h$.

Finally, we show that small errors in the core radius measurement can
induce non-linear relations between the measured collision rates and
the actual collision rates, and can, again, explain the devations
between the measured values of $\Gamma$ and the measured probabilities
that the clusters contain X-ray binaries.  Given that even the very
small measurement errors likely to be associated with the two high
quality attempts to fit King models to extragalactic globular clusters
(Jord\'an et al. 2007; Peacock et al. 2009) are likely to be
sufficient to produce spurious non-linearities, we conclude that any
fitted relation between any estimator of the collision rate and other
quantities which does not take into account both the statistical and
systematic errors on the collision rate estimates should be taken as
suspect.  We thus emphasize extreme caution in interpreting the recent
results (both based on $\Gamma_h$ measurements and on measurements of
$\Gamma$ using data with insufficient signal-to-noise) suggesting that
the relation between number density of X-ray binaries is shallower
than linear with collision rate; while of strong statistical
significance, sources of systematic error not accounted for in those
analyses can easily mimic the results found.  Typical errors in
estimations of $\Gamma$ -- even those using direct core radius
measurements in Cen~A -- are likely to have systematics such that
differences from linearity of less than 0.2 dex will be highly suspect
without detailed simulations showing that they the deviation cannot be
an artefact of asymmetric errors on the collision rate estimates.

At the present time, then, we conclude that there is no observational
evidence to contradict the idea that the only morphological parameter
of a cluster that is important for predicting the probability the
cluster will host an X-ray binary is the collision rate.  We note that
comparison of this observational finding with simple theory suggests
that the retention fraction for neutron stars in concentrated clusters
is lower than that in low concentration clusters.  This finding, if
verified, would indicate that the most concentrated clusters at the
present time were the most concentrated clusters at the time of the
supernovae producing their neutron stars.  It is thus an important
target for theory work to determine whether feasible models of the
evolution of clusters lead to such an effect.  We note that it is not
required that the clusters' concentration parameters themselves remain
fixed - just that the timescales and directions of the evolution of
the concentration parameters for different clusters are similar enough
that the retention fractions remain correlated with the present day
concentrations.

There are, of course, other parameters that clearly matter, such as
the cluster metallicity (e.g. Kundu et al. 2002).  There are also
other parameters which should matter, such as the binary fraction, and
the history of the evolution of the cluster.  However, with relatively
sparse data on the binary fractions in most clusters, it is difficult
to look for evidence that the binary fraction matters.  The best
evidence of the importance of the history of the cluster
(e.g. following Fregeau 2008) would be to find evidence of deviations
from the predictive power of $\Gamma$.  Therefore, it would make sense
to continue attempts to make good estimates of $\Gamma$ in a larger
sample of clusters.  Unfortunately, really strong progress is unlikely
to be possible until the development of 30-m class telescopes with
multiconjugate adapative optics.

\section{Acknowledgments}
We thank John Fregeau and Fred Rasio for useful discussions about the
limitations of collision rates, even in the presence of perfect data,
and Arunav Kundu and Steve Zepf for useful discussions about a variety
of topics related to optical properties of globular clusters and for
critical readings of the manuscript prior to submission.  We found the
TOPCAT package very useful in constructing some of the plots used in
this paper, and are grateful for its existence.  We are very grateful
for the report of an anonymous referee which made several suggestions
to improve the clarity of the paper, and a few suggestions for
additional work that have extended the scope of the paper
significantly and positively.

\label{lastpage}

\end{document}